\documentclass[journal,article,accept,oneauthor,pdftex,12pt,a4paper]{mdpi} 
\lastpage{x}
\doinum{10.3390/------}
\pubvolume{17}
\pubyear{2015}
\externaleditor{Academic Editor: Giorgio Kaniadakis }
\history{Received: 12 February 2015 / Accepted: 11 March 2015 / Published: xx March 2015}

\Title{Ricci Curvature, Isoperimetry and a Non-additive Entropy}

\Author{Nikos Kalogeropoulos}

\address [1]{%
Weill Cornell Medical College in Qatar, \ Education City, \ PO Box 24144, \ Doha, \ Qatar; \linebreak E-Mail: nik2011@qatar-med.cornell.edu
\vspace{-12pt}
}

\abstract{Searching for the dynamical foundations of Havrda-Charv\'{a}t/Dar\'{o}czy/\linebreak Cressie-Read/Tsallis non-additive entropy, 
we come across a covariant quantity called, alternatively, a generalized Ricci curvature, an $N$-Ricci curvature  or a Bakry-\'{E}mery-Ricci 
curvature  in the configuration/phase space of a system. 
We explore some of the implications of this tensor and its associated curvature and present a connection with the non-additive entropy under 
investigation. We present an isoperimetric interpretation of the non-extensive parameter and comment on further features of the system 
that can be probed through this~tensor.}


\usepackage{soul}

\begin{document}

\vspace{-12pt}
\section{Introduction} 

														
Havrda-Charv\'{a}t  \cite{HC} / Dar\'{o}czy  \cite{Dar} / Cressie-Read \cite{CR, CR-book} / Tsallis \cite{T1, T-book} entropy is single parameter family 
of  functionals, which have attracted some interest in the Statistical Mechanics community over the last 25 years. To go straight to the point, assume 
that a discrete set of outcomes is labelled the set~$I$.
Assume that each outcome is labelled by $i\in I$ and the corresponding probability of its occurrence is indicated by $p_i$. Then the non-additive 
entropy \ $\mathcal{S}_q$ \ that we will be interested us in this work, is defined~by 
\begin{equation}      
     \mathcal{S}_q [\{ p_i \} ] \ = \ k_B \ \frac{1}{q-1} \left\{ 1 - \sum_{i\in I} \  p_i^q \right\}  
\end{equation}
Here $k_B$ stands for the Boltzmann constant which will be set $k_B=1$ almost everywhere in the sequel. The naive extension to a
continuous set of outcomes, is characterised by the probability density function $\rho: \Omega \rightarrow \mathbf{R}$, 
and is assumed to be absolutely continuous  everywhere on \ $\Omega$ \ with respect to the Lebesgue measure (volume) \ $dvol_{\Omega}$  is
\begin{equation}      
   \mathcal{S}_q [\rho] \  = \  k_B \  \frac{1}{q-1} \left\{1 - \int_{\Omega} [\rho(x)]^q \ dvol_{\Omega} \right\}
\end{equation}
In both of the above expressions $q\in\mathbf{R}$, although a recent work has suggested \cite{WilkW} the possibility of $q\in\mathbf{C}$.
We called the definition of $\mathcal{S}_q$ for continuous sets of outcomes (2) as ``naive'' since there has been some recent controversy 
about the validity of (2) \cite{Abe1, Andr, Abe2, BOT, BoLu, LuBo, QuLi, PRo} and some related skepticism on whether it is possible to extend $
\mathcal{S}_q$  to continuous sets of outcomes. 
We consider this criticism to be valid and the general controversy not yet settled. Nevertheless, in the absence of a viable alternative or a consensus,
we will use (2) in the sequel as the version of $\mathcal{S}_q$ for a continuous sets of outcomes, having the above possible caveat in mind.
It is reassuring to notice that  for \ $q \rightarrow 1$ \ one recovers the Boltzmann/Gibbs/Shannon (BGS) entropic functional 
\begin{equation}   
    \mathcal{S}_{BGS} [\rho] \ = \ - k_B \int_{\Omega} \rho(x) \log \rho(x) \ dvol_{\Omega} 
\end{equation}


A question of fundamental importance for any entropic functional is to determine its dynamical foundations. More specifically, to determine the 
microscopic dynamical systems whose collective behaviour is encoded by the entropic functional under consideration. 
Moreover one can ask whether is it possible, even in principle for someone to   
predict which particular entropic functional, or less ambitiously, which are the common features of a class of entropic functionals that can effectively
describe the collective behaviour given a microscopic dynamical system. In our considerations, despite the fact that such an assumption can be 
considered substantially, or even unnecessarily, restrictive, we will always have in mind Hamiltonian systems of many degrees of freedom. 
The dynamical behaviour of such systems is described by their evolution in their configuration or phase space. Such a space is a manifold 
$\mathbf{M}$ endowed with a Riemannian metric $\mathbf{g}$. There may be additional structures present, such as the symplectic structure
in phase space $\mathbf{\omega}$ whose presence has unexpected and profound consequences~\cite{Gr1}. 
However in this work, we will use exclusively the underlying Riemannian manifold ($\mathbf{M, g}$) and ignore any further structures that may be 
present.

In Section 2, we discuss the basics of Ricci curvature on Riemannian manifolds and the generalized \linebreak / N- / Bakry-\'{E}mery Ricci curvature and 
the related tensor constructed via optimal transport.
In Section~3, we discuss the connection of the generalised Ricci curvature to $\mathcal{S}_q$ through a gradient flow and through the 
behaviour of the functions belonging to the displacement convexity classes $\mathcal{DC}_N$. In Section~4, we present an isoperimetric 
interpretation of the non-extensive parameter $q$. We also present an  interpretation of $q$ in terms of a projection arising from coupling the 
system to an external ``thermostat''. In Section 5, we present a brief assessment of the current situation.     

\pagebreak
 \section{About the Ricci Curvature and Its Generalizations}
\vspace{-12pt}
\subsection{Geometry in Mechanics}

Consider an autonomous dynamical system of \ $n$ \ interacting point particles, having mass \ $m_i, \ i=1,\ldots, n$. \  
Let its configuration space \ $\mathbf{M} $ \ be parametrized by the local coordinates \ $\{x^i, \dot{x}^j\}, \  i,j = 1, \ldots n$ \ and let its Lagrangian be
\begin{equation} 
                 \mathcal{L} \ =  \ \frac{1}{2} \sum_{i=1}^n m_i (\dot{x}^i)^2 - V(x^1, \ldots, x^n, \dot{x}^1, \ldots, \dot{x}^n)  
\end{equation}

The dot stands for the derivative with respect to some parameter (``time''). To make the analysis more tractable, we simplify  (4) by assuming 
generalized velocity-independent interactions, namely
\begin{equation}
V = V( \{ x^i \} ), \hspace{5mm} i = 1, \ldots, n
\end{equation}
This potential form is quite restrictive as it does not cover the case of point particles interacting with Electromagnetic or Yang-Mills fields. 
However, it turns out that  such interactions can be straightforwardly incorporated by following a similar approach
for the tangent bundle $T\mathbf{M}$ or more general, model-dependent, principal or associated bundles related to $\mathbf{M}$. 

However, it turns out that  such interactions can be straightforwardly incorporated by following a similar approach.

If we  wish to be slightly more general, by possibly incorporating the effect of constraints and reducing the system to truly independent 
variables (without constraints) even locally, we can use instead of (4) 
\begin{equation}
                \mathcal{L} \ = \ \frac{1}{2} \sum_{i,j=1}^n M_{ij} \dot{x}^i \dot{x}^j + V( \{ x^i \} ) 
\end{equation}
where \ $M_{ij}$ \ stands for a positive definite quadratic form (mass matrix) which can be used as a Riemannian metric on \ $\mathbf{M}$. \ 
In the above case (6), when all masses of particles are taken as equal to each other 
\begin{equation}
     M_{ij} = \delta_{ij}
\end{equation}
we get back (4). Since the system is autonomous, with canonical momenta \ $p_i = \delta\mathcal{L}/\delta\dot{x}^i$, \  its Hamiltonian 
\begin{equation}
     \mathcal{H} \  =  \  \frac{1}{2} \sum_{i=1}^n p_i^2 + V(x^1, \ldots, x^n)  
\end{equation}
corresponding to total energy \ $E$ \ of the system, is an integral of motion. Maupertuis' principle, recast in Hamiltonian terms, states that the 
motions of the system are the stationary paths in \ $\mathbf{M}$ \ of the adiabatic functional 
\begin{equation}                              
        \mathcal{F}[c] = \int_{c(t)} p_i \ \dot{x}^i \ dt
\end{equation}
where \ $c: [0,1] \rightarrow T^{\ast} \mathbf{M}$,  are  rectifiable curves with fixed initial and final points 
having total energy \ $\mathcal{H} = E$. \  Maupertuis' principle, as generalized by Hamilton, states that physical motions are extremals 
of the length functional (9) of such curves \ $c$. \ In geometric terms, such extrema are geodesics whose local characterisation in terms of the 
coordinate components of the metric 
\begin{equation}
     \mathbf{g}(\partial_i, \partial_j) \ = \ g_{ij}, \hspace{10mm} i,j, = 1. \ldots, n 
\end{equation}
are geodesics on the level sets of energy \ $E$ \ of  \ $\mathbf{M}$. \ 
The geodesics equation in such local coordinates  is
\begin{equation}
               \frac{d^2 x^i}{ds^2} + \Gamma^i_{\ jk} \frac{d x^j}{ds} \frac{d x^k}{ds} = 0
\end{equation}
where the Christoffel symbols (connection coefficients)  $\Gamma^i_{\ jk}$ are given, as usual, in terms of the 
metric tensor components \ $g_{ij}$ \ by 
\begin{equation}
   \Gamma^i_{\ jk} = \frac{1}{2} g^{il} \left( \partial_j g_{lk} +\partial_k g_{jl} - \partial_l g_{jk} \right)
\end{equation}
where \ $\partial_i  = \partial / \partial x^i$ \ and the Einstein summation convention is assumed over repeated indices in (9)-(12) and henceforth, unless otherwise stated.

Here, and in the sequel, \ $s$ \ represents the arc-length of a curve $c$ which is assumed to be arc-length parametrised.  
An alternative to the \ $M_{ij}$ \ metric, is the Jacobi  metric
\begin{equation}
    g_{ij} = 2 \left\{ E-V(x^1, \ldots, x^n) \right\}  M_{ij}
\end{equation}
for which the geodesic equation (11) reduces to Newton's Second Law
\begin{equation} 
      \frac{d^2 x^i}{dt^2} = - \frac{\partial V}{\partial x_i}
\end{equation}

In classical Statistical Mechanics we consider dynamical systems having sets of ``nearby'' initial conditions. Phrased differently,
we are not considering just one particular geodesic \ $c_0$ \  to describe the evolution of the system, but rather a set of them, all of which are initially 
close to \ $c_0$. \ The behavior of nearby geodesics is encoded, in the linear approximation, via the Jacobi fields with coordinate components 
$J^i, i=1, \ldots, n$. In their dynamical evolution,  the infinitesimal separation between nearby geodesic is determined by the  
by the geodesic deviation (Jacobi) equation. This is a linearization of a variation of the equation of geodesics and, 
expressed in the local coordinates we are using, it is  
\begin{equation}
     \frac{d^2 J^i}{ds^2} + R^i_{\ jkl}\frac{dx^j}{ds}J^k \frac{dx^l}{ds} = 0
\end{equation} 
where \ $R^i_{\ jkl}$ \ are the components of the Riemann tensor in the coordinate basis of \ $T\mathbf{M}$.  


\subsection{Rudiments of  Riemannian Curvature }

The Riemann tensor $R$ with components in a coordinate basis indicated by $R^i_{\ jkl}$ is a fundamental object in Riemannian geometry.
It should be noted that despite a century and a half of intense exploration, many of its properties still remain unknown \cite{Gr-SGMC}. A way to 
motivate its introduction is by searching for a local quantity that allows us to determine the distance between two points in a Riemannian 
manifold. Such a calculation is practically intractable, as can be seen by looking at the geodesic equation (11) in conjunction with (12). The 
Riemann tensor is a quantity that allows us to infinitesimally address such a question. This viewpoint, among several others, can be seen in 
\cite{Gr-SGMC}.  

The components of the Riemann tensor are explicitly given, in terms of the metric, by
\begin{equation}
       R^i_{\ jkl} = \partial_k \Gamma^i_{\ jl} - \partial_j \Gamma^i_{\ kl} + \Gamma^i_{\ km}\Gamma^m_{\ jl} - \Gamma^i_{\ jm} \Gamma^m_{\ kl}
\end{equation}

The Jacobi field with coordinate components \ $J^i, \ i= 1, \ldots, n$ \ points from \ $c_0$ \ toward nearby geodesics, as noted above,
but may also point along \ $c_0$ \ itself.
The definition of the Riemann tensor is far more elegant and brings forth its linear and differential nature when expressed in terms of the 
Levi-Civita connection $\nabla$ (the unique symmetric connection preserving the metric) as a multi-linear map 
$R: T\mathbf{M} \times T\mathbf{M} \times T\mathbf{M} \rightarrow T\mathbf{M}$     
\begin{equation}
       R(X,Y) Z \ = \ \nabla_X\nabla_Y Z - \nabla_Y \nabla_X Z - \nabla_{[X,Y]} Z
\end{equation}  
where \ $X, Y, Z \in T\mathbf{M}$ \ and $[\cdot, \cdot ]$ indicates the Lie Bracket on $T\mathbf{M}$. 

By taking advantage of  (15), $J^i$  can be chosen to be perpendicular to \ $c_0$. \ We see from (16), that the behavior of nearby 
geodesics is controlled by the Riemann curvature tensor (17). Consideration of initial conditions near \ $c_0$ \ amounts to 
averaging in the n-1 perpendicular directions to $c_0$, as expressed by \ $J^i, \ i=2,\ldots,n$, \ by using a suitably chosen 
measure. The simplest case is to choose a uniform measure, meaning a measure which is a constant multiple of the Lebesgue 
measure (``Riemannian volume'') $dvol$. Then evolution of a set of initial conditions along \ $c_0$ \ is 
controlled by the average of the Riemann tensor in the 2-planes spanned by the tangent to $c_0$ and one of the perpendicular directions to \ $c_0$, 
\ an average that results in the Ricci tensor along \ $c_0$. \ 
Therefore, the Ricci tensor is given by the following, essentially unique, contraction of the Riemann tensor

\begin{equation}
      R_{ij} = R^k_{\ ikj} 
\end{equation}

For geometric purposes, one uses the following far more transparent, but essentially equivalent as can be proved via polarisation, scalar quantities.   
Their formulation is slightly easier on orthonormal bases. So, consider an orthonormal basis of \ $T\mathbf{M}$ \ with respect to the metric 
tensor \ ${\bf g}$  \ indicated by \ $\{e_i\}, \ i=1, \ldots, n$. \  Let \ $e_1$ \ be the tangent to $c$.  
The sectional curvature in a 2-plane of $T\mathbf{M}$ at $P\in \mathbf{M}$ spanned by \ $e_p, e_q, \ p\neq q, \ p,q = 1, \ldots, n$ \ is defined by 
\begin{equation} 
     k_P (e_p, e_q) \  =  \ R_{ijkl} (e_p)^i (e_q)^j (e_p)^k (e_q)^l
\end{equation}
without a  summation over $p,q$. In a slightly different notation, following the notation of (17), we can express the sectional curvature as 
\begin{equation}
    k_P (e_p, e_q) \ = \ \mathbf{g} (R(e_p, e_q) e_q, e_p)      
\end{equation}
where no summation over $p$ or $q$ takes place, once more.

As defined, the sectional curvature is formally a function on the Grassmann manifold \ $G_{2,n} (\mathbf{M})$ \ of 2-dimensional planes of 
$T\mathbf{M}$. So, the sectional curvature is, at its core, an essentially 2-dimensional quantity. Its geometric meaning becomes evident 
through the following two theorems which are true in 2-dimensions. Let $P$ be a point of the Riemannian manifold $\mathbf{M}$ and let 
$r$ be the radius of a geodesic circle around $P$. Let $c(r)$ be its circumference and $A(r)$ be the area of the geodesic disk 
whose boundary has length  $c(r)$. Then \cite{Spivak2}
Bertrand-Puiseux (1848)       
\begin{equation}
      k_P \ = \ \lim_{r\rightarrow 0}  \  3 \cdot \frac{2\pi  r - c(r)}{\pi r^3} 
\end{equation}
Essentially equivalently \cite{Spivak2} (Diquet 1848) 
\begin{equation}
    k_P \ = \  \lim_{r\rightarrow 0}  \  12 \cdot \frac{\pi r^2 - A(r)}{\pi r^4}
\end{equation}
According to these  theorems one sees that the sectional curvature determines how much the volume of a sphere or ball exceeds that of the 
corresponding sphere or ball in flat space. There are several other, equivalent, formulations of this geometric fact \cite{CE, Sakai}.


\subsection{About the Ricci Curvature}

The Ricci curvature in the direction of \ $e_1$, \  which as was stated above is assumed to be tangent to the geodesic $c$ passing by $P$, 
is a symmetric bilinear form related to the Ricci tensor by   
\begin{equation}
     Ric(e_1, e_1) \ =  \ R_{ij}(e_1)^i(e_1)^j 
\end{equation}
In other words, it is given in terms of the sectional curvature at $P\in\mathbf{M}$, 
\begin{equation}
   Ric(e_1, e_1) \ = \ \sum_{i=2}^n \ k_P (e_1, e_i)
\end{equation}
This makes evident the fact that the Ricci curvature is the outcome of averaging in all directions perpendicular to $e_1$ at $P$ as was previously mentioned.
This averaging is also explicit in the contraction in the indices of the Riemann tensor resulting in the Ricci tensor (18). 
Hence the definition of the Ricci tensor or Ricci curvature uses explicitly two distinct facets of Riemannian manifolds: their metric (and connection
which is uniquely defined through the metric) and a measure (in this case the volume). These two concepts are uniquely inter-related 
in Riemannian manifolds exactly because such manifolds $\mathbf{M}$ are locally isometric, to first order, 
to the Euclidean space $\mathbf{R}^n$ \cite{Gr-SGMC}. This can be most easily seen via the expansion of the metric in 
geodesic normal coordinates around $P\in\mathbf{M}$ taken as the origin of the normal coordinate system 
\begin{equation}
     g_{pq} (\{ x^i \}) \ = \ \delta_{pq} + \frac{1}{3} R_{prqs} (P) \ x^r x^s + O(||x||^3)
\end{equation}
One can define the volume in Riemannian spaces via the following two requirements \cite{Gr-SGMC}   
\begin{itemize}
   \item For surjective, distance-decreasing maps  \ $f: \mathbf{M}_1 \rightarrow \mathbf{M}_2$, \  the volume obeys \   
                             $vol \ \mathbf{M}_1 \ \leq \ vol \ \mathbf{M}_2$.
   \item The volume of the unit cube \ $[0, 1]^n$ \ in \ $\mathbf{R}^n$ \ is normalized so that \ $vol \ ([0,1]^n) = 1.$                              
\end{itemize} 
and then prove the more familiar formula in local coordinates
\begin{equation}
      dvol \ = \ \sqrt{g} \ dx, \hspace{10mm} g = \det {g_{ij}}
\end{equation}

There are several ways for someone to understand the geometric meaning of Ricci curvature \cite{CE, Sakai, Gr-SGMC, Zhu}. A very 
common one, which will be useful in the sequel,  is through the Bishop-Gromov volume comparison theorem. The main idea behind comparison 
theorems \cite{CE, Sakai, Gr-SGMC, Zhu, Esch, Kar, Gr-book} is that even though it is quite hard to analyze geometric features of manifolds 
by directly solving the partial differential equations equations resulting from 
infinitesimal considerations, such as the geodesic (11) or the Jacobi equations (15) , one may still be able to extract useful geometric and topological 
information about such manifolds by comparing them to ``simpler'' spaces. As such ``model'' spaces one usually considers the 
simply-connected manifolds of constant sectional curvature (``space forms''). In this spirit, one tries to establish inequalities, bounding
the quantities of the manifold of interest by the corresponding ones of the model space. Such ideas can also be extended to more 
general metric spaces, quite frequently also endowed with measures \cite{Gr-book}. This ``comparison" idea is not foreign in Physics, 
where occasionally one encounters such inequalities, especially in more rigorous treatments of Statistical Mechanics \cite{Simon, Gal}. 

There are different versions of the Bishop-Gromov volume comparison theorem expressing the same idea slightly differently. For our purposes
we use the following: Assume that $(\mathbf{M, g})$ is a complete Riemannian manifold with $Ric \geq (n-1)k, \ k\in\mathbf{R}$ and 
$P\in\mathbf{M}$ an arbitrarily chosen point. Let $B_P(r)$ be the geodesic ball of radius $r$ centered at $P$ and $B_k(r)$ be the ball of radius $r$
in the space form of constant sectional curvature $k$ and dimension $n$. Then the function
\begin{equation}            
     r  \  \longmapsto \  \frac{vol \ B_P(r)}{vol \ B_k(r)}
\end{equation}
is a non-increasing function of $r$. This volume ratio approaches $1$  as $r \rightarrow 0$ due to (25). 
An immediate corollary of this is (Bishop's inequality)
\begin{equation} 
         vol \ B_P(r) \ \leq \  vol \ B_k(r) 
\end{equation}
From a physical (General Relativistic) viewpoint the Bishop-Gromov volume comparison theorem (27)  is ``obvious''. Let $Scal$ indicate the 
scalar curvature, namely the unique contraction of the Ricci tensor $Scal = R_{\mu\nu} g^{\mu\nu}$.  Einstein's equations 
\begin{equation}
        R_{\mu\nu} - \frac{2}{n} \ Scal \ g_{\mu\nu} \ = \ \frac{8\pi G}{c^4} \ T_{\mu\nu}
\end{equation}
determine the metric $g_{\mu\nu}$ in terms of the stress energy tensor $T_{\mu\nu}$. Here $c$ stands for the speed of light and $G$ is the 
universal gravitational constant. Roughly speaking, in areas where there is a lot of matter, $T_{\mu\nu}$ will result in strong gravity, so the bound 
in the Bishop-Gromov inequality will be high. Consider a homogeneous and isotropic space-time, 
as in the case of the Friedmann-Robertson-Walker (FRW) 
cosmology, as a comparison space. Any space-time that has more attractive matter than that, will make the geodesics converge faster,  
as seen also in (21), hence the volume of a ball centered along one such member of the geodesic congruence  will contract faster than in the 
model (FRW) space. The above analogy relies in the fact that the results we use in the Riemannian signature carry over to the Lorentzian signature 
case, something that is known, even though non-trivial to prove \cite{BEE}.  

A second implication of Ricci curvature, still in the context or Comparison Geometry, and of potential interest in Physics is via Lichnerowicz's 
inequality \cite{Lich}. This provides a lower bound on the lowest non-trivial eigenvalue $\lambda_1$ of the Laplacian $\Delta$ in terms of a lower 
bound on the Ricci curvature. As a reminder, the Laplacian $\Delta$ on functions $f: \mathbf{M} \rightarrow \mathbf{R}$ on a Riemannian manifold 
$(\mathbf{M, g})$ is defined to be the trace of the Hessian  
\begin{equation}
        \Delta \ = \ tr \ (\mathrm{Hess}) 
\end{equation}
or expressed in coordinates $\{ x^i, \ i=1,\ldots,n\} $ in terms of the components of the metric $g_{ij}$ 
\begin{equation}
      \Delta  f \ = \ - \frac{1}{\sqrt{g}}  \partial_i \left(  g^{ij} \sqrt{g} \ \partial_j  f                    \right) 
\end{equation}
where summation over repeated indices is assumed. The significance of $\lambda_1$  for Physics is 
substantial: in the particular context of Lagrangian field theories, it expresses the lowest excitation of the mass spectrum, or alternatively  
and depending on the interpretation, the mass of the lightest particle in the particle spectrum. Since it is not practically feasible to explicitly 
calculate $\lambda_1$ even for the simplest manifolds, with scant few exceptions, providing bounds to it  
is the best that someone can hope for. Lichnerowicz's theorem states that for $(\mathbf{M, g})$ a compact Riemannian manifold
without boundary with 
\begin{equation}
                  R_{ij} \ \geq  \ k g_{ij}
\end{equation} 
for $ \mathbf{R} \ni k > 0$, then $\lambda_1$ satisfies
\begin{equation}       
                \lambda_1 \  \geq  \ \frac{n}{n-1} k       
\end{equation}
It is worth observing that the right hand side is $\lambda_1$ of the sphere $\mathbf{S}^n$ endowed with 
the round metric, having constant curvature \ $k / (n-1)$. \ A result by Obata \cite{Oba} states that the equality is attained if and only if 
$(\mathbf{M, g})$ is actually isometric to such a round sphere. This result can be interpreted as stating that knowing the mass of the  lowest excitation, 
alongside the Ricci curvature lower bound (32), uniquely determines the whole spectrum of excitations in this space(-time). 
It actually determines much more than that: it completely determines the geometry and the topology of such a space(-time). 
In General Relativity, or diffeomorphism-invariant theories, the  Ricci curvature bound  (31) usually results from imposing a 
strong energy condition \cite{BEE}, namely a lower bound to 
\begin{equation}
       \left( T_{\mu\nu} - \frac{1}{n-2} \ T_{\rho}^{\rho} \ g_{\mu\nu} \right)  X^{\mu}Y^{\nu} 
\end{equation}
where $X, Y \in T\mathbf{M}$ with some additional causal behaviour assumed. Such a requirement (33) 
constrains the properties of what a  ``reasonable'' mass-energy distribution in space(-time) should be allowed to have. 
In the above paragraph, we use the word ``space(-time)''  loosely, as we always assume that $\mathbf{g}$ has a positive-definite signature.  


\subsection{Generalized Ricci Curvature}

There are several ways to go about generalising the above concepts. There is not any really unique extension of the Ricci curvature in more general 
metric-measure spaces, but several proposals, slightly different from each other exist. There are  several different motivations for such 
generalizations. The most immediately pertinent one for our purposes, is the claim that \ $\mathcal{S}_q$ \ describes systems that are non-ergodic.
As a result, in such cases one does not have available a uniqueness statement like the Krylov-Bogoliubov theorem \cite{KH}, 
or a statement  relating the asymptotic averages along a trajectory (or iterates of maps)  and phase space (micro canonical) averages like 
Birkhoff's ergodic theorem and its implications \cite{KH}. On the other hand, things may not be as uncontrollable as they may appear at first sight: 
indeed according to the Ergodic Decomposition theorem \cite{KH}, every invariant Borel probability measure of a continuous map on a metrisable 
compact space can be decomposed into a sum of ergodic invariant probability measures each of which is supported in disjoint subsets 
of the whole space. In practical terms though, concretely determining such a decomposition into sets having ergodic measures may not be tractable.
So, to proceed along these lines, we assume that a non-ergodic measure is absolutely continuous with respect to the volume of the underlying 
configuration / phase space $\mathbf{M}$  whose projection on the total energy $E$ hyper-surfaces $\mathbf{M}_E$  gives rise to the 
micro-canonical measure. Hence it is of interest to determine a generalisation of the Ricci tensor that is defined with respect to the measure
\begin{equation}          
         d\mu \ = \ e^{-f} \ dvol_\mathbf{M}
\end{equation}
rather than with respect to the volume element, where $f: \mathbf{M} \rightarrow \mathbf{R}$. So, the question is to define a Ricci-like tensor that 
will infinitesimally control the behaviour of such a $d\mu$ as in (34). The easiest approach, used extensively in Physics,  is to construct, by hand, a simple 
two-index symmetric tensor  from the scalar $f$ and from the Ricci tensor $R_{ij}$. Symmetric tensors constructed from $f$ and involving two derivatives are  \ 
$(\partial_i f)(\partial_j f)$  \ as well as the Hessian      
\begin{equation}
      \mathrm{Hess} \  f \ = \ \nabla df
\end{equation}
which for $X, Y \in T\mathbf{M}$ amounts to 
\begin{equation}
      \nabla df (X,Y) \ = \ X(Y(f)) - (\nabla_X Y) (f)   
\end{equation}
In local coordinates \ $x^i, \ i=1, \ldots, n$ \ the Hessian can be written as 
\begin{equation}
   (\mathrm{Hess} \ f)_{ij} \ = \ (\partial_i\partial_j -  \Gamma^k_{\ ij} \partial_k) f   
\end{equation}
where the summation convention upon repeated indices is assumed.
So the easiest choice would be to define a generalised Ricci tensor \ $(R_N)_{ij}$ \ by
\begin{equation}
    (R_N)_{ij} \ = \ R_{ij} + (\mathrm{Hess} \ f)_{ij}  
\end{equation}
This definition is essentially due to Bakry and \'{E}mery \cite{BakEm} where $N$ is a letter which is just part of the notation, at this stage. 
The realisation that the term $(\partial_if)(\partial_jf)$ can also be included with an arbitrary undetermined 
coefficient $N\in [1, \infty)$ and the exploration of some of its consequences was provided first by Qian \cite{Qian}.  
Therefore, a more general definition of $R_N$ is provided by 
\begin{equation}
     (R_N)_{ij} \ = \  R_{ij} + \mathrm{(Hess} \ f)_{ij}  - \frac{1}{N-n} (df \otimes df)_{ij}
\end{equation}
The exact normalisation in front of the last term is a matter of convention. What is not a matter of convention though, 
is that in this expression there is the undetermined coefficient $N$ which has to be inserted by hand since it is not
determined by any of the differential or algebraic (such as the Bianchi or the Palatini) identities that the Riemann tensor 
and its contractions obey.

 The geometric origin of $N$ is straightforward to pinpoint: in the Riemannian case, the metric uniquely determines the volume 
element and the Hausdorff dimension of the manifold expresses this unique relation between the metric and the volume, as 
in the discussion preceding (26).   In the case of a general metric-measure space, the measure is not related in any unique way to the metric. 
In the case of Finsler spaces for instance, which from a particular viewpoint are considered to be a ``direct" generalisation 
of the Riemannian manifolds, no unique measure is preferred, so several have been proposed depending on one's goals
\cite{Gr-book, AlvP}.  Therefore, the effective dimension
of the measure \ $d\mu$ \ (35) in a metric-measure space cannot be captured by the  Hausdorff dimension $n$ of 
$\mathbf{M}$, since $n$ is really expressed in terms of the volume of $\mathbf{M}$. 
Hence the effective dimension of $d\mu$ is an additional piece of information that has to be provided, {\em a priori}. 


One can put together the definitions (39) and (40) by observing that $\lim_{N\rightarrow\infty}$ of (40) reduces to (39). 
Putting all these elements together, one can state that the central differential quantity for characterising the non-ergodic 
behaviour of systems conjecturally described macroscopically by $\mathcal{S}_q$ is the generalized- / N- / Bakry-\'{E}mery- 
Ricci tensor defined by         
\begin{eqnarray}
           (R_N)_{ij} \ = &     \left\{  \begin{array}{ll}
                                                    R_{ij} + \mathrm{(Hess} \  f)_{ij},   &   \ \mathrm{if} \ \ N = \infty\\
                                                                                    &    \\
                                                   R_{ij} + \mathrm{(Hess} \ f)_{ij}  - \frac{1}{N-n} (df \otimes df)_{ij}, & \ \mathrm{if} \  \ n<N<\infty\\
                                                                                    &    \\
                                                  R_{ij} + \mathrm{(Hess} \ f)_{ij}  - \infty\cdot (df \otimes df)_{ij}, & \ \mathrm{if} \ \  N=n\\
                                                                                    &     \\
                                          -\infty,    & \ \mathrm{if} \ \ N<n     
                                          \end{array} \right.
 \end{eqnarray}
where, by convention, \ $\infty\cdot 0 = 0$. \ The last two lines are put there for completeness and for having a unifying treatment 
of these extreme cases with the rest. The definition (40) is taken from the works of J. Lott and C. Villani, in particular \cite{LV-Annals, V-book}. 
We observe that the generalized Ricci tensor $R_N$ is, in reality, a one-parameter family of tensors giving non-trivial results when 
$N \in [n, \infty ]$.  

It may not come as a total surprise that such an extension of the Ricci tensor and the related curvature is not unique, see  \cite{Ohta, Oll1, Oll2} 
for some alternatives, for instance. We will mention the synthetic definitions which reduce to (41) in the case of smooth measure spaces in the 
following Sections. The definition of (41) is a matter of choice, 
which however has many desirable properties \cite{LV-Annals, V-book}. Our interest to this  choice having some more widespread, 
or even fundamental,  significance comes from that it has been ``re-discovered" and used independently in different contexts. An example 
of the former occasion is via the work of Bakry-\'{E}mery \cite{BakEm} and Qian \cite{Qian}. The first two authors are interested in properties 
of the heat semigroup in the presence of external conservative forces which provide an additional drift term to the heat equation. Their 
analytical approach lead them to a Bochner formula whose further analysis lead to the definition of the curvature-dimension $CD(k,n)$ 
condition. Details can be found in~\cite{BakEm, Led}. This condition expresses properties of manifolds of sectional curvature at least $k$ and 
dimension at most $n$ in a way which is strongly reminiscent of the Gromov pre-compactness theorem  \cite{Gr-book}. 
A second occasion, is the case of the extensive use of the $N=\infty$ expression in (41), in particular, in the Ricci flow 
leading to the proof of the Poincar\'{e}, and consequently  of Thurston's geometrization, conjecture by G. Perelman~\cite{Per1, Per2, KlLo}.        
A third occasion was the work of Chang, Gursky and Yang \cite{ChGuY} who asked on whether one could  define conformally invariant 
analogues of the Ricci and scalar curvatures on smooth metric measure spaces. The comparison and interpolation between \cite{BakEm} 
and \cite{ChGuY}  was taken up by Case~\cite{Case1, Case2}, who has also contributed to the investigation of the quasi-Einstein metrics
$(R_N)_{ij} = \lambda g_{ij}$ \ resulting from the definition of (41) for \ $N<\infty$. \ In the Physics literature, the generalised Ricci 
curvature has been recently discussed in explorations of scalar-tensor (e.g., dilaton, Brans-Dicke {\em etc.}) theories of gravity~\cite{Wool, RupW, GalW}.

The geometric meaning of $(R_N)_{ij}$ starts becoming clearer when one tries to check on whether, or under what conditions, 
standard results of Riemannian geometry can be extended to smooth metric measure spaces \cite{Lott-H, WW, LV-Annals, V-book}. A theorem in the 
spirit of the Bishop-Gromov volume comparison goes as follows: 
Consider a compact, smooth metric measure space $(\mathbf{M, g}, \mu )$ \ where following (35),  
 \begin{equation}
 \mu (B_P(r)) \ = \ \int_{B_P(r)} \ e^{-f} \ dvol_{\mathbf{M}}
\end{equation}
having \ $(R_N)_{ij} \geq k g_{ij}, \ \ k \geq 0$  \  for some $N\in [1, \infty)$.  \  Then for all \ $0 < r \leq R$ \ and \ $P\in \mathbf{M}$ \ one has           
\begin{equation}
       \frac{\mu (B_P(R))}{\mu (B_P(r))}    \    \leq  \    \left( \frac{R}{r} \right)^N   
\end{equation}
The theorem actually holds for the class of measured length spaces, which are more general than Riemannian manifolds. Usually volume 
comparison theorems in this spirit also require some additional conditions (such as bounds, convexity properties {\em etc.}) on $f$. It is worth noticing 
that according to the Bishop-Gromov comparison theorem, the weighted volume of $\mathbf{M}$ does not expand faster than polynomially, 
with the exponent being exactly  $N$. This justifies, to some extent, the statement made above that $N$ can be seen a substitute for the 
Hausdorff, or as an effective dimension, of the measure $\mu$ on $\mathbf{M}$.   

Regarding Lichnerowicz's inequality, the generalization goes as follows: Since the measure of integration changes from \ $dvol$ \ to \ $d\mu$ \
the Laplacian has to be modified if we want to keep it a self-adjoint operator with respect to  $d\mu$. The new Laplacian is 
\begin{equation}
       \Delta_f \ = \ \Delta - \mathbf{g}(\nabla f, \nabla\cdot )  
\end{equation}
It is straightforward to check that this is indeed a self-adjoint operator with respect to the new measure $d\mu$, namely that for 
$\varphi_1, \varphi_2  \in L^2(\mathbf{M}, \mu)$, one has
\begin{equation}
    \int_{\mathbf{M}} \  \varphi_1 \Delta_f \varphi_2  \ d\mu \ = \  - \int_{\mathbf{M}} \mathbf{g} (\nabla\varphi_1, \nabla\varphi_2 ) \ d\mu \ = \ 
                              \int_{\mathbf{M}}  (\Delta_f \varphi_1) \varphi_2 \  d\mu 
\end{equation}
It should be noted at this point that, physically, this is the Laplacian on functions of $\mathbf{M}$ but in an external  field whose potential is $f$.
Therefore, maintaining the self-adjoint property of the Laplacian amounts to modifying its sub-leading symbol by adding a drift term. 
Generalizing Lichnerowicz's theorem one can prove that if \ $(R_N)_{ij} \geq k g_{ij}$, \ with \ $k > 0$,  \ then
\begin{equation} 
      \lambda_1 \ \geq \  \frac{N}{N-1} k
\end{equation}
Comparing (46) with (33) we see, once more, the validity of the interpretation of $N$ as a dimension of the measure \ $\mu$. \ It is quite 
interesting to notice that in the case of $N=\infty$ the resulting inequality is $\lambda_1 \geq k$ \ which is independent of $N$. This has 
important implications in the sequel as the $N=\infty$ case will turn out to be related to $\mathcal{S}_{BGS}$ when a finite $N$ is related to 
$\mathcal{S}_q$.    

Having stated all the above, it appears, and correctly so, that the considerations are quite generic and could be satisfied by any functional form 
related to non-ergodicity. This is true, in part.  It turns out that an infinity of entropic functionals could play a significant role, in this respect. 
The functional  $\mathcal{S}_q$ has some special 
significance even among them, although it is not unique. How these elements work together, will be discussed in the next Section.


\section{Ricci Curvature via Optimal Transport}

There are several  threads of development leading to the relation between $\mathcal{S}_q$ and the generalised Ricci 
curvature $R_N$. For a comprehensive treatment of these topics, one can start by consulting \cite{V-book}.


\subsection{Otto's View: the Porous Medium Equation and  the Geometry of Space of Probability Distributions}

An influential idea that connects $\mathcal{S}_q$ with the above developments is due to Otto \cite{Otto}. He studied the set of solutions of 
the porous medium equation in $\mathbf{R}^n$
\begin{equation}
          \partial_t \rho - \Delta \rho^m \ = \ 0   
\end{equation}
where $\rho \geq 0$ is a density function on $\mathbf{R}^n$ which explicitly depends on time $t\in [0, \infty)$. In this equation \
$m \geq \frac{n-1}{n}$ \ and \ $m > \frac{n}{n+2}$ for reasons that will  be stated  later.
The question that arose was how to interpret the porous medium equation as a gradient flow equation. 
As a reminder, a gradient flow \cite{AGS, Per1, Per2, KlLo} of sufficient generality for our purposes, consists of a Riemannian manifold 
$(\mathbf{M,g})$ and an ``energy'' functional $E [\rho ]$ on $(\mathbf{M, g})$ obeying the autonomous differential equation
\begin{equation}  
    \partial_t \rho  \ = \ - \nabla E [\rho ] 
\end{equation}
A simple example of such a gradient flow is the heat/diffusion equation where $E[\rho]$ is the kinetic energy (Dirichlet functional) of $\rho$.
The interesting property of a gradient flow is that  $E[\rho ]$ decreases along the actual trajectories $\rho(t)$. \ The established practice until 
\cite{Otto} for the porous medium equation (47), was the following: The manifold $\mathbf{M}$  of its solutions was taken to be
\begin{equation} 
    \mathbf{M} \ = \ \left\{ \rho: \mathbf{R}^n \rightarrow \mathbf{R}_+ \ \mathrm{such \  that}  \ \int_{\mathbf{R}^n} \rho \ dvol \  = \ 1 \right\}  
\end{equation}
where $dvol$ indicates the infinitesimal volume element on the space we are integrating over (in this case $\mathbf{R}^n$). \ 
Its tangent space at \ $\rho \in \mathbf{M}$,  \ $T_\rho \mathbf{M}$, \  is
\begin{equation}
    T_\rho \mathbf{M} \ = \ \left\{ s: \mathbf{R}^n \rightarrow \mathbf{R}, \ \mathrm{such \ that} \ \int_{\mathbf{R}^n} s \  dvol \ = \ 0 \ \right\}  
\end{equation}
The tangent space \ $T_\rho \mathbf{M}$ \ can also be seen as 
\begin{equation} 
   T_\rho \mathbf{M} \ = \ \left\{ \phi : \mathbf{R}^n \rightarrow \mathbf{R} \right\} / \sim 
\end{equation}
where $\sim$ identified $\phi$ differing by an additive constant and where  $\phi$ were solutions of the Poisson equation
\begin{equation} 
    - \Delta \phi \ = \ s
\end{equation}
The metric tensor \ $\mathbf{g}$ \ at \ $\rho \in \mathbf{M}$ \ was taken to be 
\begin{equation}
   \mathbf{g}_\rho (\phi_1, \phi_2) \ = \ \int_{\mathbf{R}^n} (\nabla \phi_1 )(\nabla \phi_2) \ dvol  
\end{equation}
The ``energy'' functional was taken to be 
\begin{equation}
     E [\rho ] \ = \ \frac{1}{m+1} \  \int_{\mathbf{R}^n} \rho^{m+1} \ dvol
\end{equation}
Otto's view was to keep (49)--(51) as they were, but change (52) to requiring $\phi$ to satisfy
\begin{equation}
       -\nabla \cdot (\rho\nabla\phi ) \ = \ s 
\end{equation}
and modify the metric $\mathbf{g}$ on $\mathbf{M}$ from (53) to
\begin{equation}
     \mathbf{g}_\rho (\phi_1, \phi_2) \ = \ \int_{\mathbf{R}^n} \  (\nabla\phi_1)(\nabla\phi_2) \ \rho \ dvol
\end{equation}
One can immediately see from (49), (53), (56) that ($\mathbf{M, g}$) is an infinite dimensional Riemannian manifold, so appropriate care 
should be taken of convergence issues in a more careful, less formal, treatment.  
The ``energy'' functional was modified from (54), by Otto, to be
\begin{equation}
    E[\rho] \ = \ \left\{
\begin{array}{lll}
      \frac{1}{n-1} \int_{\mathbf{R}^n} \rho^m  \ dvol & \mathrm{ for} & m \neq 1\\
               & & \\                      
      \int_{\mathbf{R}^n} \rho \ \log\rho \ dvol    & \mathrm{for} & m = 1
\end{array}
        \right.
\end{equation} 
We recognise right away that the ``energy'' functional in (57) has the exact same form as the entropy $\mathcal{S}_q$ if $q$ is identified with $m$.
We also see that $\mathcal{S}_{BGS}$ plays the role of the ``energy'' functional in the gradient flow characterisation of the ordinary heat/diffusion 
equation which corresponds to $m=1$ in~(47).
Moreover we see the prominent role that the modified measure $\rho \ dvol$, which is exactly (35) with $\rho = \exp (-f)$, plays in the viewpoint 
advocated in \cite{Otto}. Otto went much further along these lines. He established a formal calculus on $(\mathbf{M, g})$, dubbed ``Otto calculus'',
without paying too much attention to the subtleties arising from the fact that ($\mathbf{M, g}$) is infinite-dimensional. He discovered, for instance, that 
the manifold ($\mathbf{M, g}$) has formally non-negative sectional curvature. Addressing more carefully some of 
these topics was undertaken by Lott \cite{Lott-GCWS}, among many others \cite{Pet, LV-Annals, Ohta2, Lott-TCWS}  and several aspects are still under 
intense investigation.

The long-term asymptotic behaviour of the porous medium equation is expressed by the Barenblatt solution $\rho_B$. 
To determine the scaling behaviour of a solution in the asymptotic regime close to $\rho_B$, we re-express it as   
\begin{equation}
   \rho (t, x^i ) \ = \ \frac{1}{t ^{n\alpha}}  \  \tilde{\rho} \left( \log t,  \frac{x^i}{t^\alpha}  \right), \hspace{10mm}     i=1,\ldots, n  
 \end{equation}
where 
\begin{equation}
    \alpha \ = \ \frac{1}{2+n(m-1)}     
\end{equation}
One can prove that $\tilde{\rho}$ satisfies the gradient flow equation, where \  $\tau = \log t$
\begin{equation}
    \partial_\tau \tilde{\rho} \ = \ - \nabla \tilde{E}[\tilde{\rho}]
\end{equation}
where the modified energy functional $\tilde{E}$ is given by 
\begin{equation}
    \tilde{E} [\tilde{\rho}] \ = \ E[\tilde{\rho}] + \alpha V[\tilde{\rho}]
\end{equation}
and where $V$ is represents the second moment functional of $\tilde{\rho}$
\begin{equation}
  V[\tilde{\rho}] \ = \ \frac{1}{2} \int_{\mathbf{R}^n} |x|^2 \tilde{\rho}(t, x^i) \ dvol
\end{equation}
with $|x|^2$ being the Euclidean norm of $\{ x^i, \ i=1,\ldots n \} \in \mathbf{R}^n$. Otto proved \cite{Otto} that $\tilde{E}$ is uniformly 
strictly convex on  ($\mathbf{M, g}$) since it satisfies 
\begin{equation} 
     \mathrm{Hess} \ \tilde{E} [\tilde{\rho}] \ \geq \ \alpha \mathbf{1},  \hspace{5mm}  \forall \ \tilde{\rho} \in \mathbf{M}
\end{equation}
and moreover that 
\begin{equation}
     \frac{d}{d\tau} \left( e^{2\alpha\tau} \nabla\tilde{E} [\tilde{\rho}]  \right) \ \leq \ 0 
\end{equation}

An interpretation of these results in the context of $\mathcal{S}_q$ is as follows: First of all, the convergence to the Barenblatt solution, 
which we have not explicitly stated, but whose scaling is the same as that of (58) can be traced back to the origins of $\mathcal{S}_q$
in the Physics literature \cite{T1, T-book}: fractals usually do have some form of self-similarity which is expressed by scaling properties.
Second: the ad hoc constraints on the value of $m$ stated above, can now be seen as 
requirements for the convexity of $E$ and for the well-posedness and finiteness at $\rho_B$  of $E$ and $V$.
Third, we observe in (64) that the rate of convergence of $\rho$ to $\rho_B$ is polynomial (power-law) in terms of $t$.
This, conjecturally, is one of the important properties of systems described by $\mathcal{S}_q$: their power law rate of convergence to 
their long-time asymptotic limits \cite{T-book}. It is worth mentioning that this rate of convergence is  believed to also have the form of a $q$-exponential      
\begin{equation}
     e_q(x) \ \equiv \ \{ 1+ (1-q) x\}^\frac{1}{1-q},  \hspace{5mm} 1\neq q\in\mathbf{R} 
\end{equation}
However, the value of the non-extensive parameter controlling the rate of approach to equilibrium in the $q$ exponent distribution is 
not necessarily the same as that in the equilibrium distribution itself \cite{T-book}. 
There is no {\em a priori} reason why these two distributions  that share the same functional form should also have the same non-extensive parameter. 
In the absence of any physical arguments to the contrary, we would expect that different aspects of the systems under consideration are 
described by different values of the non-extensive parameter, if we can even assume
that their macroscopic behaviour is described by $\mathcal{S}_q$ in the first place. Moreover, if the renormalisation group lessons 
are taken at heart, it is also entirely possible that $q$, for a given system, may be different for the approach to different quasi-equilibrium states.  


Leaving these possibilities on the side, we can also see in (64) an aspect related to the existence of the observed, in numerical simulations, 
quasi-equilibrium states \cite{AB-Book, ABB, Zas}. If the approach to equilibrium is polynomial, then it is too slow, with respect to an exponential time 
scale that we usually employ in analyzing the approach of systems to equilibrium. This way of thinking is similar to what we advocated in \cite{NK1, NK2},
for justifying why the largest Lyapunov exponent of systems described by $\mathcal{S}_q$ is zero. In closing, we would like  to refer to 
\cite{Vaz-Book} for a comprehensive treatment of the porous medium equation, including the Barenblatt solution stated above, 
and \cite{Vaz-Frac} for properties of some of its possible fractional generalisations.


\subsection{Optimal Transportation and Wasserstein Spaces} 

In a seemingly different line of development from the above, the definition of the generalized Ricci curvature and its relation to $\mathcal{S}_q$ 
resulted from attempts to address aspects of the Monge optimal transport problem. This problem ascribed to Monge \cite{Monge}  
tries to determine the least costly way of transporting 
a quantity of soil/earth (``d\'{e}blais'') from a pile to an excavation/embankment (``remblais''). Naturally, one has to specify before-hand 
what is the cost function $c$ which is used to determine such a total cost. Although the problem was initially posed in $\mathbf{R}^n$, it can clearly be 
formulated for general complete separable metric (Polish) spaces endowed with Borel probably measures.
It may be worth mentioning right away, that the general problem remains unsolved even for simple cost functions in  $\mathbf{R}^n$, 
although substantial  progress has been made in special cases, especially during the last two decades.

The most obvious cost function $c$ with a geometric significance, is the Euclidean distance $d(x, y)$ between of the points $x$ 
in the pile and points $y$ in the 
excavation. This choice of cost function $c(x,y)$ turned out to be difficult to deal with, the root cause being its lack of uniform convexity properties.
To get a simpler problem without deviating too much from the original, one uses instead as cost function
\begin{equation}              
     c(x,y) \ = \ \frac{1}{2} \ d^2(x,y)
\end{equation}
which turns out to be more manageable as it is uniformly convex in $\mathbf{R}^n \times \mathbf{R}^n$.

Monge's transport  problem can be mathematically 
formulated as follows \cite{AmbG, Evans, LV-Annals, McC-G, V-book} (these are general references for  this whole subsection):
 consider two Borel measures $\mu^+, \mu^-$ on $\mathbf{R}^n$,  or in more general Polish measure spaces mentioned above, such that 
\begin{equation}  
      \mu^+ (\mathbf{R}^n) \ =  \ \mu^-( \mathbf{R}^n) 
\end{equation}
both of which are finite. In addition, consider the class of Borel maps $f: \mathbf{R}^n \rightarrow \mathbf{R}^n$  pushing 
forward $\mu^+$  to  $\mu^-$  namely
\begin{equation}
     (f_\sharp \mu^+) (A) \     =     \  \mu^+ (f^{-1} A), \hspace{5mm} \forall  \ A \subset \mathbf{R}^n
\end{equation}
This can be translated as requiring 
\begin{equation}
     \int_{A_1} w(f(x)) d\mu^+(x) \ = \ \int_{A_2} w(y) d\mu^-(y) 
\end{equation}
for all continuous functions $w: \mathbf{R}^n \rightarrow \mathbf{R}$ \ with $A_1, \ A_2 \subset \mathbf{R}^n$ 
indicating  the support of \ $\mu^+, \ \mu^-$ \ respectively.  Let the set of such functions be indicated by $\mathcal{F}$. 
Given a cost function $c: \mathbf{R}^n \times \mathbf{R}^n \rightarrow [0, \infty)$  consider the total cost functional 
\begin{equation}
     I_M [f] \ = \ \int_{\mathbf{R}^n} c(x, f(x)) d\mu^+ (x) 
\end{equation} 
The goal is to determine the existence and find the properties of an optimal transport $f_o \in \mathcal{F}$ such that
\begin{equation}
        I_M [f_o] \ = \ \min_{f \in\mathcal{F}} \ I_M [f]
\end{equation}
This problem has proved to be quite difficult to address. There are at least three reasons for this
\begin{itemize} 
    \item The problem is highly non-linear. To see this more concretely, let's assume that both $\mu^+, \mu^-$ are absolutely continuous with respect to the 
               volume element of $\mathbf{R}^n$ with corresponding Radon-Nikodym densities $\rho^+, \rho^-$. Then it turns out that the
               push forward condition (68) translates into the Monge-Amp\'{e}re, non-linear, equation 
                      \begin{equation}
                                 \rho^+ (x) \ = \ \rho^- (f(x)) \ \det (\nabla f)
                      \end{equation} 
               where $\det$ stands for the Jacobian determinant of the differential map $\nabla f$.        
   \item  Such a solution may not exist: consider for instance $\mu^+$ to be the Dirac delta function but not  $\mu^-$.
   \item The transport condition (69)  is not weakly sequentially closed and this creates the additional complication 
             that the minimum (71) may not be realized, as subsequences of approximating maps may not converge in any 
            reasonably weak topology.                        
\end{itemize}

Progress toward solving Monge's problem was slow until Kantorovich \cite{Kant1, Kant2} substantially reformulated it in more amenable terms.
The difficulties associated with  the Monge problem are partly attributable to the fact that the optimal transport map the $f_o$ that we 
seek cannot ``split'' the measure $d\mu^+$. Heuristically, $f\in\mathcal{F}$ cannot move a part of $d\mu^+$  in a location of $\mathbf{R}^n$
and another part in some other location of $\mathbf{R}^n$, as such a map would not be well-defined. Such a constraint (68) is too strong 
to handle. Instead, Kantorovich made the following two modifications. 
First, he transformed the problem into a linear one, by recasting it as follows: Consider the space $\mathcal{P}(\mathbf{R}^n  \times \mathbf{R}^n)$ 
of Borel probability measures $\mu$ whose push-forwards $p_1, p_2$ on the first and second $\mathbf{R}^n$ factors respectively 
are $\mu^+$ and $\mu^-$  
\begin{equation}  
     \mathcal{P} \ = \ \{  \mu : \mathrm{Probability \ measures \ on} 
                 \ \mathbf{R}^n \times \mathbf{R}^n \  \  \mathrm{such \ that} \ \  (p_1)_\sharp \ \mu \ = \ \mu^+, \ \  (p_2)_\sharp \ \mu \ = \  \mu^- \} 
\end{equation}
The elements of $\mathcal{P}$ are called transference plans. Consider then, for \ $\mu \in \mathcal{P}$ \ the modified cost functional 
\begin{equation}
      I_K [\mu ] \ = \ \int_{\mathbf{R}^n \times \mathbf{R}^n} c(x,y) \ d\mu (x,y)
\end{equation}
This functional is linear in $\mu$ so if some assumptions on the cost function can be made, compactness can ascertain the existence of at least 
one minimizing measure, called optimal transference plan, \ $\mu_o\in\mathcal{P}$
\begin{equation}  
      I_K [\mu_o ] \ = \ \min_{\mu \in \mathcal{P}} \  I_K [\mu ]
\end{equation}
It should be stressed at this point that there is no {\em a priori} guarantee that an optimal transference plan $\mu_o$ of the Kantorovich functional $I_K$ (75)
corresponds to  a mapping  $f_o$ sought after in Monge's functional $I_M$ (71).


The second crucial observation allowing to make progress in solving
the Kantorovich problem is the Kantorovich - Koopmans \cite{Kant1, Kant2, Koop} duality: in the spirit of duality in convex analysis and geometry 
they reformulated the variational  problem (74) so that instead of a minimisation, it became a maximisation problem. To formulate this dual variational 
problem, define the set
\begin{equation}    
     \mathcal{M} \ = \  \{ (\phi, \psi): \mathbf{R}^n \times \mathbf{R}^n \rightarrow \mathbf{R} \ \ \mathrm{continuous}, \ \phi (x) + \psi (y) \ \leq \ c(x,y) \} 
\end{equation}
Then, introduce the dual of the Kantorovich functional \ $I_{K^\ast}$ \ as 
\begin{equation}
    I_{K^\ast} [\phi, \psi ] \ = \ \int_{\mathbf{R}^n} \phi(x) \ d\mu^+ (x) \ + \ \int_{\mathbf{R}^n} \psi(y) \ d\mu^- (y)  
\end{equation}
The Kantorovich-Koopmans duality is the statement that the minimisation problem (75) is equivalent to the maximisation 
\begin{equation} 
         I_{K^\ast} [\phi_o, \psi_o ] \ = \ \max_{(\phi, \psi) \in \mathcal{M}} I_{K^\ast} [\phi, \psi] 
\end{equation}
This problem is amenable to the methods of linear programming and as such it is easier to solve than the original (70). What is still not clear 
though is whether a solution of  (78) can be derived from a solution of~ (71). The converse is clearly true: suppose that $f_o$ is 
a minimizer of (71). Then $\mathrm{Id} \times f_o: \mathbf{R}^n \rightarrow \mathbf{R}^n \times \mathbf{R}^n$ \ and \ 
$\mu \ = \ (\mathrm{Id} \times f_o)_\sharp \mu^+ \in \mathcal{P}$ \ minimizes \ $I_K$ in (74) so it is a solution to (75).  

Let's be more general in this paragraph and assume that $\mathbf{M}$ is a Polish space (complete and separable metric space) endowed with a 
distance function $d$ and let $\mu \in\mathcal{P}(\mathbf{M} \times \mathbf{M})$ be a transference plan with marginals $\mu^+$ and $\mu^-$ as above. 
Moreover, let's assume that $\mathbf{M}$ is compact, for simplicity. Let's choose as cost function 
\begin{equation} 
      c(x,y) \ = \ d^p(x,y), \hspace{5mm} p\geq 1
\end{equation}
The Kantorovich functional (74) with such a cost becomes 
\begin{equation} 
        I_K [\mu ] \ = \ \int_{\mathbf{M} \times\mathbf{M}} d^p(x,y) d\mu(x,y)
\end{equation}
Then (80)  can be considered as a measure of the discrepancy between the marginals $\mu^+, \mu^-$. 
Naturally, there are many ways to define the discrepancy between measures  \cite{Rachev}, depending on one's goals.  
For our purposes, the optimal transference plan (75) with the cost function (79), in other words the minimizer of (80), 
enters naturally in the picture.
It is interesting to notice that such an optimal transference plan always exists under the above assumptions \cite{V-book} so the 
variational problem (75) does have an actual solution. If, in addition one sets  
\begin{equation} 
      \{ W_p (\mu^+, \mu^-)\} ^p \ = \ I_K[\mu_o] 
\end{equation}
or, in other words, if one defines for any two Borel measures $\mu_1, \mu_2 \in \mathcal{P}(\mathbf{M} \times \mathbf{M})$  
\begin{equation}
     W_p (\mu_1, \mu_2) \ = \  \left( \min_{\mu \in \mathcal{P}} \ \int_{\mathbf{M}\times\mathbf{M}} d^p(x,y) \ d\mu (\mu_1, \mu_2) \right)^\frac{1}{p}
\end{equation}
where the minimum is taken over all possible transference plans $\mu\in \mathcal{P}$ with marginals $\mu_1, \mu_2$, then it can be 
immediately checked that $W_p (\mu_1, \mu_2)$  satisfies all the properties of a distance function. As such, it is called the 
$p$- Monge-Kantorovich-Rubinstein-Vasherstein distance, or in short and even though it is a partial misnomer, the $p$-Wasserstein distance (metric).
All these metrics, except $W_\infty$, give the same topology on a compact $\mathbf{M}$ which is the weak-$\ast$ topology. 
For such a distance function to be non-trivial the integral in (82) must converge. This becomes quite important especially when $\mathbf{M}$ 
is non-compact. Hence, the definition of the $p$-Wasserstein distance 
restricts the elements of $\mathcal{P}$ to only the ones with finite $p$-moments. The number of converging moments required for each case of $p$  
is the only essential difference between the different $W_p$ 
metrics on $\mathcal{P}$. The subspace $\mathcal{P}_p (\mathbf{M}\times\mathbf{M})$ 
of  $\mathcal{P}(\mathbf{M}\times\mathbf{M})$ of measures with converging $p$-moments,  
endowed with the Wasserstein distance then becomes  a metric space with several desirable 
properties \cite{V-book} and is called the $p$-Wasserstein space of $\mathbf{M}$. We forego expanding on this topic of current research as 
further information is not needed in the subsequent discussion.. 


\subsection{The Brenier Map and Its Extensions: the Role of Convexity} 

We go back to $\mathbf{R}^n$, initially at least, and to assuming that $c(x,y) = \frac{1}{2} d^2(x,y)$ 
where $d$ denotes the Euclidean distance in $\mathbf{R}^n$.
We would like to determine a solution to Monge's problem in this case, namely to minimize
\begin{equation} 
    I_M[f] \ = \  \frac{1}{2} \int_{\mathbf{R}^n} d^2(x, f(x)) \ d\mu^+(x)
\end{equation}
Skipping a detailed explanation of the reasons why \cite{Bren1, Bren2}, we just state the result  \cite{McCann} (Brenier map):  
Let $\mu^+$ and $\mu^-$ be probability measures on $\mathbf{R}^n$ which are absolutely continuous with respect to the Lebesgue measure (volume) 
$vol$ of $\mathbf{R}^n$. Then there exists a convex function $\varphi: \mathbf{R}^n \rightarrow \mathbf{R}$ whose gradient 
$f = \nabla\varphi: \mathbf{R}^n \rightarrow \mathbf{R}^n$ pushes forward $\mu^+$ to $\mu^-$. This map is unique, almost everywhere, and it 
therefore provides a unique solution to Monge's problem (70). Immediate generalizations are due to~\cite{GangMc}. In the case of the Brenier map,
the Monge-Amp\'{e}re equation (72) for the corresponding densities $\rho^+, \rho^-$  takes the more familiar form
\begin{equation}  
     \rho^+ (x) \ = \ \rho^- (\nabla\varphi (x)) \  \det (\nabla^2\varphi(x))
\end{equation}
as long as $\nabla\varphi (x)\in \mathbf{R}^n$. The convexity of $\varphi$ implies that Jacobian  obeys $\det (\nabla^2\varphi) \geq 0$
and it also guarantees that $f$ is almost everywhere differentiable. To prove this, one can use the dual Kantorovich formulation (77).  
The constraint in (76) becomes
\begin{equation} 
     \phi(x) + \psi(y) \ \leq \ \frac{1}{2} d^2(x,y)
\end{equation}
Now, changing variables to 
\begin{equation}
     u(x) \ \equiv \ \frac{|x|^2}{2} - \phi (x), \hspace{15mm} v(y) \ \equiv \ \frac{|y|^2}{2} - \psi (y)
\end{equation}
(85) amounts to 
\begin{equation}
       u(x) + v(y) \ \geq x\cdot y
\end{equation}
so, (78) reduces to minimizing the functional 
\begin{equation}
      \tilde{I}_{K^\ast} [u,v]  \ = \ \int_{\mathbf{R}^n} u(x) \ d\mu^+(x) \ + \ \int_{\mathbf{R}^n} v(y) \ d\mu^-(y)
\end{equation}
under the constraint (87). By standard results of convex conjugacy theory \cite{Rock}, we know that  
there is such a pair $u_o, v_o$ minimizing (88) and moreover these two functions are 
Legendre-Fenchel  transforms of each other, namely they obey
\begin{equation}
     u_o(x) \ = \ \max_{y\in\mathbf{R}^n} \{ x\cdot y - v_o(y) \},   \hspace{15mm}  v_o(y) \ = \ \max_{x\in\mathbf{R}^n} \{ x\cdot y - u_o(x) \}    
\end{equation}
We will skip the non-trivial topic of the regularity of solutions to these variational problems altogether, referring to  \cite{V-book, McC-G} for 
some statements and an extensive list of references on such issues.

The question that is raised now is whether the above results can be extended from $\mathbf{R}^n$ to Riemannian manifolds ($\mathbf{M, g}$). 
This was addressed in part by \cite{OV, McC2, CEMcS1} \ and by \cite{CEMcS2} for metric measure spaces. One modification that needs to be 
made is to replace the Legendre-Fenchel transform (86) by the generalized convexity transform 
\begin{equation}
      h^c (y) \ = \ \inf_{x\in\mathbf{M}} \{ c(y,x) - h(x) \},   \hspace{10mm} h^{cc} (x) \ = \ \inf_{y\in\mathbf{M}} \{ c(x,y) - h^c (y) \}
 \end{equation}
for $h: \mathbf{M} \rightarrow \mathbf{R} \cup \infty$ with the cost function $c(x,y)$ having the usual form $c(x,y) = \frac{1}{2} d^2(x,y)$ as for 
$\mathbf{R}^n$. Then Brenier's theorem \cite{McC2} is extended as follows. For a closed, connected Riemmanian manifold ($\mathbf{M, g}$)
let $\mu^+$ be absolutely continuous with respect to $vol_\mathbf{M}$ and $\mu^-$ arbitrary. Then there is a map 
$\varphi: \mathbf{M} \rightarrow \mathbf{R}$ such that $\varphi^{cc} = \varphi$ ($\varphi$ is a $c$-concave function) so that the map 
$\exp (-\nabla\varphi )$ pushes forward $\mu^+$ to $\mu^-$. Such a map is a unique minimizer within the set 
$\mathcal{P} (\mathbf{M}\times\mathbf{M})$ for Monge's transportation functional (83). In the language of the 2-Wasserstein space 
$W_2(\mathbf{M})$,  the Monge transport between $\mu^+, \mu^-$ takes place along the unique Wasserstein geodesic joining them.   
 The last statement is substantially non-trivial, because if $\mathbf{M}$ were a general length space, there could  be an uncountably 
 infinite number of geodesics joining the two measures. 
 
 A second modification addresses the issue of convexity on Riemannian manifolds. There is no unique extension of convexity from $\mathbf{R}^n$ 
 to a Riemannian manifold $\mathbf{M}$. Even then, one can either work directly geometrically with the space at hand, utilising properties of the 
 distance or the volume functions, or express convexity indirectly through functional inequalities such as the Brunn-Minkowski, the Brascamp-Lieb, 
 the Prekopa-Leindler {\em etc.} In \cite{CEMcS1, CEMcS2} the second approach was taken. Results in Comparison Geometry     
 were derived and extensively used. We will state just one such result to pave the way to the synthetic definition of the generalised Ricci curvature below.
 To generalise the linear interpolation $tx+(1-t)y$ between two points $x, y \in \mathbf{R}^n$ to the case of $\mathbf{M}$ with distance function $d$,
  they~\cite{CEMcS1} start by considering the set
 \begin{equation}
        Z_t (x,y) \ = \ \{ z\in\mathbf{M}, \ \mathrm{such \ that} \ \ d(x,z) \ = \ t d(x,y) \  \ \mathrm{and} \ \ d(y,z) \ = \ (1-t) d(x,y)  \}       
 \end{equation}
 where $t\in [0,1]$  and for $z$ lying between $x,y$.  This definition can be extended from a point $y$ to the ``target'' set $Y \subset \mathbf{M}$ as
 \begin{equation}
       Z_t (x, Y) \ = \ \bigcup_{y\in Y} Z_t (x,y)
 \end{equation}
 If $P\in\mathbf{M}$, consider the open ball centered at $P$ of radius $r$ by $B_P(r)$, and introduce a volume distortion by the volume ratio
 \begin{equation}
     w_t (x, P) \ = \ \lim_{r\rightarrow 0} \ \frac{vol \ Z_t (x, B_P(r))}{vol \ B_P (tr)}
 \end{equation}
 for $t\neq 0$. Obviously $w_1 (x,P) = 1$, and $w_t (x,P) \geq 1$ if the curvature is non-negative. whereas the opposite is true if the curvature is 
 non-positive. In $\mathbf{R}^n$ clearly $w_t(x,P) = 1$. 
An interpretation of (93) from a relativistic viewpoint, pretending for a moment that we are working in a space of Lorentzian signature, 
since light does travel on null geodesics  $w_0(x,P)$ represents the magnification of the area of a small light source located near $P$ 
as is seen by an observer at point $x$.  We compare this distortion 
 function (93) with that of the standard $n$-dimensional space forms having constant sectional curvatures $k>0$ and $k<0$ 
 respectively. The Ricci curvature in all these case is $Ric = (n-1)k$. Then for $k\in\mathbf{R}^n$ we set 
 \begin{equation}  
    sn_k (t) \ = \ \left\{
         \begin{array}{cc}
              \frac{\sin \sqrt{k} d}{\sqrt{k} d}, & k>0\\ 
                          1,   &  k=0\\
              \frac{\sinh \sqrt{k} d}{\sqrt{k} d}, & k<0\\             
         \end{array}
             \right. 
 \end{equation}
 Then, if ($\mathbf{M, g}$) is such that $Ric \geq (n-1)k$ along any geodesic of length $d(x,y)$ one has from  the Bishop-Gromov comparison theorem (27) 
 \begin{equation}
        w_t(x,y) \ \geq \  \left\{ \frac{sn_k (td(x,y)) }{sn_k (d(x,y))} \right\}^{n-1} 
 \end{equation}
 where equality holds when $\mathbf{M}$ has constant sectional curvature $k$. 
 This result be used in the sequel in the synthetic definition of the generalised Ricci curvature.
 
 As in Otto's work for $\mathbf{R}^n$, it was noticed that providing lower bounds on the Hessian (63) which is related to the energy functional (57), 
 could be useful in determining rates of convergence of gradient flows on $\mathbf{M}$ to their asymptotic configurations. For the ``energy'' 
 functional (57) with $m=1$ \cite{OV} calculated its formal Hessian of $\mathcal{P}(\mathbf{M})$ and found that it is bounded from below 
 by $k \mathbf{g}_\rho$ (56)  as long as the generalized Ricci tensor $(R_N)_{ij}$ for $N=\infty$ of $\mathbf{M}$ is bounded from below by 
 $k g_{ij}$. Hence  the k-convexity, as in (63), of an appropriate energy functional in $\mathcal{P}(\mathbf{M})$ is intimately related to the 
 a lower bound of the generalized Ricci curvature of $\mathbf{M}$.  \ The converse  statement, still using the $m=1$ case of the ``energy'' functional
 in (57),  was established by \cite{vonR-St}. The obvious question is whether these considerations can be extended to the cases of the ``energy'' functional in                   
 (57) for $m\neq 1$ and how would this influence the definitions of the generalized Ricci curvature. An answer was provided by 
 \cite{Sturm1, Sturm2, Sturm3, Sturm4, LV-Annals, LV-WCC}. 


\subsection{Displacement Convexity and Synthetic Definition of the Generalized Ricci Curvature}  

In this subsection we follow \cite{LV-Annals, LV-WCC} and can be even more general than before by considering, not necessarily smooth, 
metric measure spaces ($\mathbf{X}, d, \nu$). Here $\nu$ is Borel will be considered as a ``reference measure''. For Riemannian 
manifolds $\mathbf{M}$ the role of $\nu$ is played by the volume, but a generic metric measure space lacks such a natural choice. 
It turns out that without substantial loss of generality, the constructions remain largely the same if we only  consider measures  
$\mu$ absolutely continuous with respect to $\nu$ with Radon-Nikodym density $\rho$, namely
$\mu = \rho \nu$. The general case is discussed in \cite{LV-Annals, LV-WCC}. The first step is to consider a function 
$U: [0, \infty) \rightarrow \mathbf{R}$ which is continuous and convex. Second, given $\mu, \nu \in\mathcal{P}(\mathbf{X})$, 
define the functional  $U_{\nu} (\mu)$ by
\begin{equation}
       U_{\nu} (\mu) \ = \ \int_\mathbf{X} U(\rho (x)) d\nu(x) 
\end{equation}
The role of these definitions is to allow us to construct ``energy functionals'' for a  gradient flow akin to (57).  
Actually, from a physical viewpoint these are the entropy functionals $\mathcal{S}_q$ and $\mathcal{S}_{BGS}$ in the first and second 
lines of (57) respectively. Hence the original assumptions on the function $U$. To recover (57), consider the single-parameter family of 
functions $U_N: [0, \infty) \rightarrow \mathbf{R}$ with $N\in [1, \infty]$, which is given by 
\begin{equation}  
     U_N(r) \ = \  \left\{ 
             \begin{array}{ll}
                  Nr \left(1 - \frac{1}{r^N} \right), & \mathrm{if} \ \ N\in (1, \infty)\\
                 r \log r,  & \mathrm{if} \ \ N = \infty 
             \end{array}  
          \right.               
\end{equation}
Then 
\begin{equation}
    U_{N, \nu} \ = \  \left\{
           \begin{array}{ll}
                 N - N \int_\mathbf{X} \rho^{1 - \frac{1}{N}} \ d\nu, & \mathrm{if} \ \ N\in (1,\infty) \\ 
                 \int_\mathbf{X} \rho \log \rho \ d\nu,  & \mathrm{if} \ \ N = \infty 
           \end{array}
              \right.
\end{equation} 
Upon setting 
\begin{equation}
         N \ = \ \frac{1}{1-q}
\end{equation}
in (98), we recover (57), or equivalently (2), (3). This also fits nicely with a standing assumption of our prior work such as \cite{NK3, NK4, NK5, NK6}  
in which we have examined properties of $\mathcal{S}_q$ for values of the non-extensive parameter $q\in [0, 1]$ which are equivalent to $N\in [1, \infty].$\\

At this point, one can ask what additional properties should the functions $U$ obey so as to qualify as being the foundation of the functionals 
(96) which should be ``reasonable''. Since functionals such as (96) can be interpreted as entropies, one can search for desirable properties that 
thermodynamic potentials possess \cite{Ruelle, Simon, Gal, T-book} and demand these to hold in (96) or equivalently for the functions $U$. 
This motivated the introduction of the displacement convexity classes $\mathcal{DC}_N$ and $\mathcal{DC}_\infty$  by \cite{McC-DC}. 
A continuous, convex function belongs to the displacement convexity class $\mathcal{DC}_N$, {\em i.e.}, $U\in \mathcal{DC}_N$, for $N\in [1, \infty)$, if 
\begin{equation}  
         u(t) \ = \ t^N \ U \left( \frac{1}{t^N} \right)
\end{equation}
is convex on $(0, \infty)$. Moreover, a continuous, convex function belongs to $\mathcal{DC}_\infty$, {\em i.e.}, $U\in \mathcal{DC}_\infty $, if 
\begin{equation} 
     u(t) \ = \ e^t \ U(e^{-t}) 
\end{equation}
is convex on $( -\infty, +\infty)$. We can see that, if $N \leq N'$, then $\mathcal{DC}_{N'} \subset \mathcal{DC}_N$. More importantly, for our purposes, 
consider the following two functions that can be derived from $U$ if it is sufficiently smooth, which have the form of Legendre transforms
\begin{equation} 
    p(r) \ = \ r \ \frac{dU(r)}{dr} - U(r) 
\end{equation}
which is motivated by the definition of the pressure, if $U$ were to be the internal energy, at the level of functions rather than functionals.
Moreover consider the ``iterated pressure''
\begin{equation}
      p_2(r) \ = \ r  \ \frac{dp(r)}{dr} - p(r)
\end{equation}
Then, with the above assumptions on $U$ and also assuming that it is smooth, we have that $U\in \mathcal{DC}_N$ if and only if the function
\begin{equation} 
       r \ \longmapsto \ \frac{p(r)}{r^{1 - \frac{1}{N}}} 
\end{equation}
is non-decreasing on $(0, \infty )$. Equivalently $U\in\mathcal{DC}_N$ if and only if 
\begin{equation}
       p_2(r) \ \geq   \  - \frac{p(r)}{N}
\end{equation}
Among them, (104) shows the role of the function (97) giving rise to (98) and equivalently (57), (2)  in $\mathcal{DC}_N$. This function (97) 
is by no means unique. However it has the best possible asymptotic behaviour that preserves the  convexity property (100) defining 
$\mathcal{DC}_N$. Something similar can be said about the  class $\mathcal{DC}_\infty$: if $U\in\mathcal{DC}_\infty$, then either $U$ is linear 
or there exist $a,b >0$ so that $U(r) \geq a r \log r - br$. The latter functional form results to $\mathcal{S}_{BGS}$ for a probability density $\rho$. 
From the present viewpoint therefore, the function (97) giving rise to (98)  
is not unique at all, but has a very desirable functional form for membership in $\mathcal{DC}_N$ and $\mathcal{DC}_\infty$ that moreover 
gives rise to the entropic functionals (2), (3).

As  was also hinted in the last paragraph of Section 3.4, the definition of lower Ricci curvature  bounds for a metric measure space 
is expressed via bounds on the convexity properties of functionals arising from functions belonging to $\mathcal{DC}_N$ or $\mathcal{DC}_\infty$,    
such as (97). The ``amount of convexity'', an example of which is shown in (63), is encoded in the definition of  $k$-convexity. 
Convexity of a smooth function $f$ on the unit interval $t\in [0,1]$, which will be assumed to parametrize a geodesic in ($\mathbf{X, d, \nu}$),  
with $x,y \in \mathbf{X}$ can be determined either by using the local criterion \ \ $\frac{d^2f}{dt^2} \geq 0$ \ \ or by  Jensen's inequality
\begin{equation}    
    f(tx+(1-ty)) \  \  \leq \ \  t f(x) + (1-t) f(y) 
\end{equation}
By the same token, for $k$-convex functions, one can either use the criterion $\frac{d^2 f}{dt^2} \ \geq \ k$ or the inequality
\begin{equation}
     f (tx+(1-t)y) \ \ \leq \ \ t f(x) + (1-t) f(y) - \frac{1}{2} \ k \ t \ (1-t) \ d^2 (x,y)
\end{equation}
Relying on these, and for $k\in\mathbf{R}$, a functional $U_\nu$ with background measure $\nu$ is called 
\begin{itemize} 
    \item $k$--displacement convex if  for any two $\mu_0, \mu_1 \ \in \ \mathcal{P}(\mathbf{X})$ and for all Wasserstein geodesics
                $\{ \mu_t \}, \ t \in [0,1]$, we have 
                       \begin{equation}
                           U_\nu (\mu_t ) \  \ \leq \ \ t U_\nu (\mu_1) + (1-t) U_\nu (\mu_0) - \frac{1}{2} k t (1-t) W^2_2 (\mu_0, \mu_1)
                      \end{equation}
           for all $t\in [0,1]$ 
    \item  weakly $k$-displacement convex, if for all $\mu_0, \mu_1 \in \mathcal{P}(\mathbf{X})$ there is at least one Wasserstein 
                    geodesic along which  (108) holds.           
\end{itemize}
For the rest of this section we follow very closely \cite{LV-Annals, LV-WCC}. 
For any $k\in\mathbf{R}$, define $\Lambda: \mathcal{DC}_\infty \ \rightarrow \ \mathbf{R}\cup \{ - \infty \}$ by  
\begin{equation}
     \Lambda (U) \  = \ \left\{
          \begin{array}{ll}
                k  \lim_{r \rightarrow 0} \frac{p(r)}{r}, & \mathrm{if} \ \ k>0\\
                                0, & \mathrm{if} \ \ k=0\\
                k \lim_{r \rightarrow \infty} \frac{p(r)}{r}, & \mathrm{if}\ \  k<0
          \end{array}
                          \right. 
\end{equation}
We say that for such a $k\in\mathbf{R}$, the metric measure space $(\mathbf{X}, d, \nu)$ has $N=\infty$ generalized Ricci curvature bounded 
from below by $k$ if the Wasserstein space $W_2(\mathbf{X})$ is weakly $\Lambda(U)$-displacement convex, namely
\begin{equation}      
    U_\nu (\mu_t) \ \leq \ t U_\nu (\mu_1) + (1-t) U_\nu (\mu_0) - \frac{1}{2} \Lambda(U) t(1-t) W^2_2 (\mu_0, \mu_1) 
\end{equation}
for any two measures $\mu_0, \mu_1 \in \mathcal{P}(\mathbf{X})$, $U\in\mathcal{DC}_\infty$ and all $t\in [0,1]$.
For the finite $N$ case we need a few more definitions: given again $k\in\mathbf{R}$ and for $N=1$, define
\begin{equation} 
        \beta_t (x_0, x_1) \ = \ \left\{
             \begin{array}{ll}
                   \infty, & \mathrm{if} \ \ k>0\\
                    1, & \mathrm{if} \ \ k\leq 0
             \end{array}
                  \right.                         
\end{equation}
for $N\in (1, \infty]$ and
\begin{equation}
     \alpha \ = \ \sqrt{\frac{|k|}{N-1}} \ \ d(x_0, x_1)
\end{equation}
In addition, define
\begin{equation}
    \beta_t (x_0, x_1) \ = \  \left\{
           \begin{array}{ll}
                \exp \left( \frac{1}{6} k (1-t^2) d^2(x_0, x_1)  \right),  & \mathrm{if} \ \ N = \infty \\       
                \infty, & \mathrm{if} \ \ 
                N<\infty, \ k>0, \ \alpha > \pi \\
                \left( \frac{\sin (t\alpha)}{t\sin\alpha} \right)^{N-1}, & \mathrm{if} \ \ N<\infty, \ k>0, \ \alpha \in [0,\pi ]\\
                1, & \mathrm{if} \ \ N<\infty, \ k=0\\
                \left( \frac{\sinh (t\alpha)}{t\sinh\alpha} \right)^{N-1},   & \mathrm{if} \ \ N<\infty, \ k<0  
            \end{array} 
                      \right.
\end{equation}
It should be noted that the last three entries of (113) are the same as the right-hand-side of (95) from which they originate.
Now consider a transference plan $\eta \in\mathcal{P}(\mathbf{X}\times\mathbf{X})$ and decompose it in terms of its marginals 
$\mu_0$ and $\mu_1$  
\begin{equation}
         d\eta (x_0, x_1) \ = \ d\eta (x_1|x_0) \ d\mu_0(x_0) \ = \ d\eta (x_0|x_1) \ d\mu_1(x_1)
\end{equation}
where $d\eta(x_0|x_1), d\eta(x_1|x_0)$ indicate the corresponding ``conditional'' measures. 
Then, the metric measure space ($\mathbf{X}, d, \nu$) has $N$-Ricci curvature bounded below by $k$, if there is some optimal transference plan 
$\eta$ from $\mu_0$ to $\mu_1$ with Wasserstein geodesic $\mu_t$, so that for all $U\in\mathcal{DC}_N$ and for all $t\in[0,1]$ 
\begin{eqnarray}  
  U_\nu (\mu_t) \ \  \leq   &  (1-t) \int_{\mathbf{X}\times\mathbf{X}} \frac{\beta_{1-t}(x_0, x_1)}{\rho_0(x_0)}\ \  
                                                   U\left( \frac{\rho_0(x_0)}{\beta_{1-t}(x_0,x_1)} \right) \  d\eta(x_0,x_1) \nonumber \\  
                  &  + \  t \  \int_{\mathbf{X}\times\mathbf{X}} \frac{\beta_t(x_0,x_1)}{\rho_1(x_1)}  \ \ U\left( \frac{\rho_1(x_1)}{\beta_t(x_0,x_1)} \right)
                                         \ \ d\eta(x_0,x_1)
\end{eqnarray}
Even though this is a far from trivial statement, one can see it as a weight-averaged version of (107) with weights provided by measure ratios akin to the 
right hand side of (95)  embedded into  the entropy functionals whose convexity properties in $W_2(\mathbf{M})$ are used to reflect the Ricci 
curvature properties of $\mathbf{M}$, all in a comparison sense. 
We observe that the  definition (115) is synthetic: nowhere have we required the metric measure space to be smooth. Its apparent drawback 
for applications to Physics is that this definition of  Ricci curvature is only defined in a comparison sense. However, it is directly related to the 
convexity properties of $\mathcal{S}_q$ in the Wasserstein space $W_2(\mathbf{X})$. On the other hand, the definition (41) of the generalised 
Ricci tensor/curvature in smooth metric measure spaces is local, so it is easy to compute, in principle, but  it appears to have nothing to do 
with $\mathcal{S}_q$. However (41) and (115) give the same result, in a comparison sense, for a measured length space, hence for a Riemannian 
manifold: The Riemannian manifold ($\mathbf{M, g}$) has its generalised Ricci curvature bounded below by $k$ (115) if and only if \cite{LV-WCC} 
its $N$-Ricci tensor (41) is also bounded below by $k$, namely $(R_N)_{ij} \geq kg_{ij}$. \ Before closing this Section, it may be worth noticing that 
for non-branching spaces such as Riemannian manifolds,  there is no real distinction between an element of the displacement convexity class
$\mathcal{DC}_N$ and (97), so in this sense the entropic functional $\mathcal{S}_q$ itself  is unique in the determination of the generalised 
Ricci curvature (115).         

 
 \section{Isoperimetric Interpretation of the Non-extensive Parameter and Related Matters} 

 We can  look closer at (99) which provides a relation between the free parameter $N$ in the definition of the generalized Ricci tensor (41)
 and the non-extensive parameter $q$ in the functional (2) under the additional tacit restriction to $q\in [0,1]$. Inverting (99) we get
 \begin{equation}
        q \ = \ \frac{N-1}{N}
 \end{equation}
 This points out to an interpretation of $q$ through $N$: $N$ can be interpreted as an effective isoperimetric dimension of the measure $\mu$
 on the smooth metric measure space ($\mathbf{M, g}, \mu$). It should be noted that the isoperimetric inequalities \cite {Gr-book, Chav} 
 (``why is a soap bubble spherical''), may arguably be the most important and influential among all the geometric inequalities \cite{BuZa}. For 
 this reason, such an interpretation of $q$ allows us to use the very extensive set of techniques and results that are known on this 
 topic, mostly in a comparison sense, in reaching conclusions that may be of physical interest. It may be worth mentioning at this point that the 
 isoperimetric dimension is an asymptotic  property of a space and can be larger than its Hausdorff dimension \cite{Chav}, even if such a  space is a 
 Riemannian manifold: consider for instance the hyperbolic $n$-space whose Hausdorff dimension is $n$, but whose isoperimetric 
 dimension is infinite, as volumes and areas of the boundaries of subspaces of hyperbolic spaces are related by linear isoperimetric inequalities.     

 However, it becomes immediately obvious, that such an interpretation has a problem: indeed, even if we confine our attention 
 to $q\in [0,1]$, it seems through, mostly, data fittings for finding $q$ that the corresponding $N$ is quite small, and certainly finite, for most systems.   
 After all $q$ is a thermodynamic parameter, so it is to be expected that its value in actual systems is finite, or even ``relatively'' small.  
 On the other hand, for  Hamiltonian systems of many degrees of freedom, their configuration and phase spaces have dimensions of order of magnitude 
 $n \sim 10^{23}$  which represents a gross mismatch with the values of $N$ computed through (116) and from data fittings of $q$. We have not 
 been able to resolve this interpretational conundrum in an acceptably convincing way. The best that we can currently state, is to vaguely suggest 
 that $N$ should actually represent an effective isoperimetric dimension of the configuration or phase space per degree of freedom, 
 something more akin to $\frac{N}{n}$ in our notation. Given this, it might be of some interest to attempt to explain the appearance of the escort distributions 
 \cite{BeckS, T-book} used in the calculations of the thermodynamic parameters \cite{T-book}, under this light.      
   
The Bishop-Gromov generalised comparison theorem (43) can also provide quantitative support for the relation between (41) and (2). 
It may be worth noticing, that according to \cite{HanTh1, HanTh2}, $\mathcal{S}_q$ describes the cases of systems whose  configuration or 
phase space volume increases in a power-law, rather than an exponential, manner. Obviously checking whether this is true explicitly,  
in specific examples (of Hamiltonian systems of many degrees of freedom) seems to be quite hard, if feasible at all. However the generalized 
Ricci curvature (41) provides a feasible, at least in principle, way of doing so by using the Bishop-Gromov comparison result (43): if  $R_N >0$ then, 
according to (43), the corresponding measures on the configuration / phase space increase slower than a power-law fashion with exponent 
equal to $N$. This also provides another dimensional interpretation of $N$ as the maximal Hausdorff dimension exponent bounding the expansion 
of measures in the configuration or phase space of the microscopic~system.
  
There is  a slightly different way of  understanding the effect of lower bounds of (41); it is in a way, a geometric analogue to the construction of the 
canonical ensemble in equilibrium Statistical Mechanics. In this treatment, proposed in \cite{Lott-H}, one initially considers the system of interest, let's 
call its configuration / phase space by $\mathbf{B}$ coupled to a ``thermostat'' $\mathbf{F}$ which can be large but not infinite ($\mathbf{F}$ is 
compact). Let the combined system's configuration/phase space be indicated by $\mathbf{M}$. 
Naturally, the evolution in $\mathbf{M}$  is Hamiltonian along geodesics chosen with respect to the metric of $\mathbf{M}$. 
Project this evolution on $\mathbf{M}$ ``down" to the system of interest $\mathbf{B}$ assuming that the tangents to the geodesic curves on 
$\mathbf{M}$ and $\mathbf{B}$ have equal lengths with respect to the respective metrics. This requirement essentially amounts to assuming that the 
average kinetic energy per degree of freedom of $\mathbf{M}$, which might be called ``temperature" but it is far from obvious that it would have a physical 
meaning for generic systems,  is the same as that of $\mathbf{B}$. Then this projection $p: \mathbf{M} \rightarrow \mathbf{B}$ is actually a 
Riemannian submersion \cite{ON}. Moreover, assume that  the pushforward $p_\sharp$ of $dvol_\mathbf{M}$ is just a multiple of $dvol_\mathbf{B}$ and
let $N = \dim \mathbf{F}$.  Let the Ricci curvatures of $\mathbf{M}, \mathbf{B}$ be indicated by superscripts with respect to their corresponding metrics.
Then for any $k\in\mathbf{R}$, \cite{Lott-H} proves that if $Ric^\mathbf{M} \geq k$, then $Ric^\mathbf{B}_N \geq k$. \ So, a way to understand the meaning of 
the generalized Ricci tensor is to see it as the Ricci tensor due to the submersion of a higher dimensional space preserving the measure of the base up to a 
multiplicative constant. This statement may also make the generalised Ricci tensor quite useful for applications in theories involving higher (greater than 4) 
dimensional space-times. The result and the general ideas are also  close to the treatment of \cite{Alm, AMAA} who expressed the non-extensive 
parameter $q$ in terms of the scaling properties of the Hamiltonians of the ``thermostat'' $\mathbf{F}$ and of the system under study $\mathbf{B}$. 

There is the also the obvious question of how would one go about dealing with cases of systems having $q>1$. Whether Hamiltonian systems of
many degrees of freedom can be described by $\mathcal{S}_q, \ q \in \mathbf{R}$ was taken for granted for a considerable amount of time. 
More recently though, some dissenting opinions have appeared (see for instance \cite{LuBo, PRo}). We  will not go into this very important matter which 
deserves a thorough investigation, in the present work. Let's assume for argument's sake that the ``conventional wisdom'' is true and 
$\mathcal{S}_q$ can describe such Hamiltonian systems. Then a possible way around the restriction $q\in [0,1]$ that we use in this work may come from 
``dualities'' of the non-extensive parameter. It has been observed for a while \cite{T-book}, that systems described by $\mathcal{S}_q$ seem to have 
particularly nice properties under the following transformations/``dualities" of the non-extensive parameter
\begin{equation}       
         q \  \longmapsto \  2-q, \hspace{15mm}  q \ \longmapsto  \ \frac{1}{q} 
\end{equation}
These are a set of generators of conformal/M\"{o}bius transformations on the complex plane, assuming that $q\in\mathbf{C}$. Their existence may be 
allow us to extend the validity of elements of the formalism presented here outside the restricted range $q\in [0,1]$ that we have assumed. Whether or to what 
extent  such a goal  can actually be achieved, even formally, its implications and the origin of the dualities (117) is the subject of one of our current 
investigations \cite{NK7}.    

 
 \section{Assessment and Omissions}
 
 In this work, we have attempted to bring to the attention of our audience the existence, construction and geometric significance of the 
 generalised (Bakry-\'{E}mery-) Ricci tensor (41) as an analytic tool for probing the microscopic dynamics on the configuration or phase space for 
 systems whose collective behaviour is described by the Havrda-Charv\'{a}t/Dar\'{o}czy/Cressie-Read/Tsallis entropy (2). Such a straightforwardly 
 computable geometric quantity may be of some interest in performing concrete calculations in specific models, calulcations that may shed some light 
 in the dynamical aspects \cite{Pin} of the underlying systems of many degrees of freedom \cite{CPC} described by such functionals. Our view is that this
 generalised Ricci tensor may do for systems described by $\mathcal{S}_q$ what the ordinary Ricci tensor has done and can still do for 
 $\mathcal{S}_{BGS}$ \cite{Pin, CPC}. The generalized Ricci curvature, which has a purely synthetic definition (115) without resorting to differential properties, 
 may also be of some interest to the Quantum Gravity community for formulating the Einstein equations in a synthetic rather than in a differential way,
 since it is widely believed that differentiability and smoothness,  of space-time should be an emergent, rather than an {\em a priori} assumed, property.  
 
 
 We would like to point out that there is very little, if any at all, new material in this work. Far more extensive and authoritative treatments can be found in
 the literature, such as \cite{V-book}. Our goal was to help motivate and make somewhat more familiar, from a physical viewpoint, some ideas that lie behind 
 the construction and properties of the generalised Ricci curvature. We think that bringing such an object to the attention of the practitioners 
 of ``non-extensive'' entropy may create some interest which may eventually help elucidate, analytically, issues pertaining to the underlying dynamics of  
 systems described by $\mathcal{S}_q$. In this spirit, we have omitted entirely from the present work any discussion whatsoever about the regularity 
 and the convergence (usually in the measured Gromov-Hausdorff sense) of the underlying structures. Such important issues, and of potential physical 
 relevance, can be found in \cite{V-book},  in the original papers or in recent reviews on these topics.                                  
                                  
 
 \section{Acknowledgments}
 
 We would like to thank the referees for their careful reading of the manuscript and for their constructive criticism.  
 We are grateful to the organisers of SigmaPhi 2014  which took place on 7--11 July 2014 in Rhodes, Greece, and in particular to  G. Kaniadakis, 
 for their invitation to present a talk in the Conference, a substantial  expansion of which forms the content of the present manuscript.   
 
 \conflictofinterests{Conflicts of Interest}
 
The author declares no conflict of interest.  

                                                                                                                         

\bibliographystyle{mdpi}
\makeatletter
\renewcommand\@biblabel[1]{#1. }
\makeatother


\end{document}